# On the Coupling of Matter and Gravity to the Boundary of Space-Time[1]

A. N. Jourjine

*Center for Theoretical Physics, Laboratory for Nuclear Science and Department of Physics,*
*Massachusetts Institute of Technology, Cambridge, MA 02139, USA*

## Abstract

We show that under certain boundary conditions on the matter fields and on the fluctuations of the background metric the gravity-matter system can be coupled to the boundary of space-time through the stress-energy tensor. The connection of the formalism developed to the Casimir effect is discussed.

[1] This work was supported in part through funds provided by the US Department of Energy (DOE) under contract DE-AC02-76ERO3069.

# 1. Introduction

Relativistic extended objects - membranes and strings - with the "surface" action proportional to the "area" the object "sweeps" during its motion, have been studied extensively from classical and quantum points of view [1]. One of such objects is the boundary of space-time. The boundaries of vacuum phases were treated dynamically in the bag models of hadrons [2], the boundary of space-time as a dynamical object was considered in [3], where a vacuum decay mechanism which involves a spontaneous creation and expansion of the boundary was discussed. Finally, since the discovery of the Casimir effect, much work has been done on the boundary effects in quantum field theory [4].

However, in the work cited above the motion of the boundary-like objects is either predetermined or the coupling between the matter and the boundary is non-dynamical. Furthermore, the Casimir forces which one obtains by varying the total renormalized vacuum energy can only serve as a non-relativistic approximation of the back reaction of the matter fields on the boundary. The boundary of space-time with gravity has not been discussed as a dynamical object in the literature so far. It is interesting, therefore, to derive covariant equations of motion for the boundary from the action principle and to see whether the Casimir forces appear in the non-relativistic limit. In this letter we treat the boundary of $D$-dimensional space-time as a $D$-1 dimensional membrane and show that there exists a natural coupling of the boundary to the matter-gravity system.

# 2. Equations of Motion.

Our conventions are: $X$ denotes a $D$-dimensional manifold with Lorentzian metric $g_{\mu\nu}$ which has positive signature; $\partial X$ is the boundary of $X$ ; $Z^\mu(y^a)$ are the local coordinates of $\partial X$ in $X$ ; $n^\mu$ is the outward pointing space-like normal

$$n^\mu = \lambda \varepsilon_{\mu\nu\cdots\rho} \varepsilon^{a\cdots b} \partial_a Z^\nu \cdots \partial_b Z^\rho ,$$

where $\lambda$ is chosen such that $n^\mu n_\mu = 1$;

$$h_{ij} = \partial_i Z^\mu \cdots \partial_j Z^\nu g_{\mu\nu},$$

is the induced metric on $\partial X$ ; $h_\mu^\nu = \delta_\mu^\nu - n_\mu n^\nu$ is the projection operator from the tensor bundle of $X$ on that of $\partial X$ ; $\chi_{\mu\nu} = D_\rho n_\mu h_\nu^\rho$ is the second fundamental form. Greek indices run 1 to $D = \dim X$, while Latin ones 1 r o $D$ - 1. The coordinates $x^\mu$ always refer to $X$ and $y^a$ to $\partial X$. With our conventions

$$\int dx g^{1/2} D_\mu A^\mu = \int dy h^{1/2} n_\mu A^\mu .$$

All quantities are made dimensionless by a choice of units. All fundamental constants are put to one.

First consider an arbitrary matter system on a manifold with a boundary and a fixed background metric. The action for the system is

$$S = S_M + S_{\partial X}, \qquad (1)$$

$$S_M = \int dV \, \mathcal{L}(\phi_A), \qquad dV = g^{1/2} dx, \qquad (2)$$

$$S_{\partial X} = a \int dS, \qquad dS = h^{1/2} dy, \qquad a = const. \qquad (3)$$

Before deriving the equations of motion let us consider coordinate invariance of the action (1). The requirement that (1) is invariant with respect to the transformations that map $\partial X$ onto $\partial X$ identically: $y^a \to \tilde{y}^a = y^a$, gives us the usual covariant conservation law for the stress-energy tensor

$$D_\mu T^{\mu\nu} = 0, \qquad (4)$$

where by definition $T^{\mu\nu} = g^{-1/2} \delta S_M / \delta g_{\mu\nu}$. If, however, under the coordinate transformation $\partial X$ maps into $\partial X$ not identically: $y^a \to \tilde{y}^a \neq y^a$, the situation is slightly different since, in general, additional boundary terms will appear in $\delta S$ due to the non-vanishing of the boundary variations of fields and metric

$$\delta g_{\mu\nu} = L_\xi g_{\mu\nu}, \qquad (5)$$

$$\delta \phi_A = L_\xi \phi_A. \qquad (6)$$

Here the index $A$ combines external and internal indices and $L_\xi$ is the Lie derivative with respect to the vector field $\xi^\mu$ which is generated by an infinitesimal coordinate transformation.

Indeed a vector field $\xi^\mu$ which is non-zero at the boundary and is tangential to it $-\xi^\mu n_\mu \big|_{\partial X} = 0 -$ still induces a coordinate transformation according to (5, 6). In this case, performing such a transformation, we obtain

$$\delta S_M = \int dV \left\{ \left[ -D_\mu (\partial \mathcal{L}/\partial(\partial_\mu \phi_A)) + \partial \mathcal{L}/\partial \phi_A \right] L_\xi \phi_A + D^\nu T_{\nu\mu} \xi^\mu \right\}$$

$$+ \int dS \left\{ (\partial \mathcal{L}/\partial(\partial_\mu g_{\lambda\nu})) L_\xi g_{\lambda\nu} + \partial \mathcal{L}/\partial(\partial_\mu \phi_A) L_\xi \phi_A \right\} n_\mu \qquad (7)$$

$$+ \int dS \, n^\mu T_{\mu\nu} \xi^\nu,$$

and separately

$$\delta S_{\partial X} = a \int dS\, h^{\mu\nu} D_\mu \xi_\nu. \tag{8}$$

Note that additional terms will appear in (7) if the action (1) depends on higher than first derivatives of the metric or the *n*-bein. Using the equations of motion for $\phi_A$ and (4) we obtain that we must have

$$\int dS\, n_\mu \{(\partial \mathcal{L}/\partial(\partial_\mu g_{\lambda\nu})) L_\xi g_{\lambda\nu} + (\partial \mathcal{L}/\partial(\partial_\mu \phi_A)) L_\xi \phi_A\} + \int dS\, n^\mu T_{\mu\nu} \xi^\nu = 0, \tag{9}$$

and

$$\int dS\, h^{\mu\nu} D_\mu \xi_\nu = 0, \tag{10}$$

when $\xi^\mu n_\mu\big|_{\partial X} = 0$.

We now require that the boundary conditions on the matter fields cancel identically the first integral in the right hand side of (9) for all $\xi^\mu = h^\mu_\rho \tilde{\xi}^\rho$. Then for the matter fields we have

$$n^\mu T_{\mu\nu} h^\nu_\rho = 0. \tag{11}$$

The condition (10) is trivially satisfied, since $\int dS\, h^{\mu\nu} D_\mu \xi_\nu = \int dS\, \chi\, h^\mu \xi_\mu$, where $\chi = h^{\mu\nu} \chi_{\mu\nu}$. Thus, under the boundary conditions we required, (11) supplements the condition (4). For a static boundary and Minkowski metric on *X*, (11) takes the form

$$n_i T_{i0} = 0, \qquad n_i T_{ij} = (n_i T_{ik} n_k) n_j, \qquad n_\mu = (0, n_i). \tag{12}$$

This just means that there is no momentum flow through the boundary and that the force exerted on the boundary always points along the normal.

The main justification for the restriction on possible boundary conditions comes from consideration of equations of motion for the boundary. These are obtained by varying the range of integration in (2) and varying (3) with respect to $Z^\mu(y^a)$. However, if our theory is generally covariant then the range variation is equivalent to the variation (5, 6) such that $\xi^\mu n_\mu\big|_{\partial X} \neq 0$. Indeed in this case the variation of the volume is, for instance,

$$\delta V = \delta \int dx\, g^{1/2} = \int dx\, g^{1/2} D_\mu \xi^\mu = \int dy\, h^{1/2} h^\mu \xi_\mu,$$

i.e. the same result as we would obtain by the range variation with $\delta Z^\mu(y^a) = -\xi_\mu\big|_{\partial X}$.

The transformations (5, 6) are coordinate transformations for any internal point of X, thus only surface terms could appear in variation of all quantities.

Taking into account the boundary conditions and (4) we obtain for the boundary

$$a\chi n^{\mu} + T^{\mu\nu}n_{\nu} = 0, \qquad (13)$$

or

$$-a\chi = T^{\mu\nu}n_{\nu}n_{\nu} = 0. \qquad (14)$$

We see that the restriction on the boundary condition makes the equation of motion to have a particularly simple form. It means that the normal-normal component of the stress-energy tensor provides the source term for the "free" boundary, described by the equation $\chi = 0$.

The stress-energy tensor in (14) is symmetric. This is due to the general covariance of the action (1). Relativistic covariance implies the canonical stress-energy tensor in (14) and, in general, different dynamics for the boundary.

For the electromagnetic field the right hand side of (14) reduces to the familiar Casimir pressure in the case of the static boundary $R \times S^2$. As noted in [5] for the ideal conductor boundary - but, in fact, valid for any - the pressure on the $R \times S^2$ boundary given by the principle of virtual work coincides with that obtained from the normal-normal component of the renormalized stress-energy tensor.

There is nothing exotic about the restriction we imposed on the boundary conditions. Indeed in flat space-time for the canonical scalar field $\phi$ it is equivalent to the Neumann boundary condition. In this case $\mathcal{L} = 1/2 \, g^{1/2} \partial_{\mu}\phi \partial^{\mu}\phi - V(\phi, \cdots)$ and the boundary condition is

$$n_{\mu} \partial \mathcal{L}/\partial(\partial_{\mu}\phi)\big|_{\partial X} = n_{\mu}\partial^{\mu}\phi\big|_{\partial X} = 0. \qquad (15)$$

For the gauge field with $\mathcal{L} = -1/4 \, g^{1/2} F_{\mu\nu}F^{\mu\nu}$ it is satisfied by the bag boundary condition

$$n_{\mu} \partial \mathcal{L}/\partial(\partial_{\mu}A_{\nu})\big|_{\partial X} = n_{\mu}F^{\mu\nu}\big|_{\partial X} = 0. \qquad (16)$$

The case of spin 1/2 fields is somewhat more complicated since the corresponding Lagrangian depends on *n*-bein derivatives (in the generalization of the tetrad formalism). For spinors

$$\mathcal{L} = \frac{i}{2} g^{1/2}\left[\overline{\psi}\vec{\nabla}\psi - V(\overline{\psi},\psi,\cdots)\right], \qquad \vec{\nabla} = \gamma^{\mu}\nabla_{\mu},$$

where $\nabla_{\mu}$ is the covariant spinor derivative with respect to *n*-bein. After simple calculations we find that the well-known boundary condition

$$i\gamma_{\mu}n^{\mu}\psi = \psi, \qquad (17)$$

satisfies the restriction we require.

We now consider inclusion of gravity into the system. It is well known [6] that in the presence of the boundary the gravity action must be modified to cancel surface terms that contain normal derivatives of the variation of the metric $g_{\mu\nu}$. The simplest way to make the modification is to take for the gravity action [6]

$$S_G = \frac{1}{\kappa}\int dV\, R - \frac{2}{\kappa}\int dS\, \chi, \tag{18}$$

where $\kappa$ is Newton's constant and $R$ is the curvature scalar. Variation of the action (18) vanishes when the Einstein equations of motion are satisfied and when variation of metric vanishes on $\partial X$. $\left(\chi^{\mu\nu} - h^{\mu\nu}\chi\right)\delta g_{\mu\nu}\big|_{\partial X} = 0$.

We consider this as a boundary condition on the metric fluctuations, but not on the metric itself. This interpretation is in accordance with the approach taken in ref. [7]. We thus arrive to the total action for the system

$$S = S_G + S_M + S_{\partial X}, \tag{19}$$

Variation of the action (19) with respect to the boundary gives

$$\delta S = \int dV\left(R_{\mu\nu} - \frac{1}{2}Rg_{\mu\nu} - \kappa T_{\mu\nu}\right)\delta g^{\mu\nu} + \frac{1}{\kappa}\int dS\left(\chi_{\mu\nu} - \chi h_{\mu\nu}\right)\delta g^{\mu\nu} - a\int dS\, \chi\, n^\mu \xi_\nu,$$

where $\delta g_{\mu\nu} = D_\mu \xi_\nu + D_\nu \xi_\mu$. Using Bianci identity and (4) and when $\xi_\mu = n_\mu F(y)$ we obtain the boundary equation of motion

$$\chi^{\mu\nu}\chi_{\mu\nu} - \chi^2 - \frac{1}{2}a\kappa\chi = n^\mu n^\mu\left(R_{\mu\nu} - \frac{1}{2}Rg_{\mu\nu} - \kappa T_{\mu\nu}\right), \tag{20}$$

where we used the unique extension of $n_\mu$ off $\partial X$ in its neighborhood in $X$ defined by $n_\lambda D^\lambda n_\mu = 0$. In the case when $\xi^\mu = h^\mu_\rho \tilde{\xi}^\rho$ the Codacci equation ensures $\delta S = 0$. When the Einstein equations are satisfied, it appears that in this equation the effects of matter are cancelled by the effects of gravity. Remarkably, this is not true. The equation

$$\chi^{\mu\nu}\chi_{\mu\nu} - \chi^2 - \frac{1}{2}a\kappa\chi = 0, \tag{21}$$

does contain the stress-energy tensor. Indeed, using the expression for the curvature scalar $R'$ of the induced metric

$$R' = R - 2R_{\mu\nu}n^\mu n^\nu + \chi^{\mu\nu}\chi_{\mu\nu} - \chi^2 \tag{22}$$

and using Einstein's equations once more we obtain

$$R' + 2\kappa T_{\mu\nu}n^\mu n^\nu = \frac{1}{2}a\kappa\chi. \tag{23}$$

Eq. (23) contains only one free parameter: the "surface tension" constant a. It is tempting, in view of the fact that $R'$ contains only second derivatives of the small fluctuations of $Z^\mu(y^a)$, to remove the ambiguity by putting a = 0. Then for the boundary without surface tension we would have

$$R' + 2\kappa T_{\mu\nu}n^\mu n^\nu = 0. \tag{24}$$

Since $\kappa$ is very small we see that the strong coupling of (14) gave way for the weak gravitationally induced coupling. When $T_{\mu\nu} = 0$ then $R' = 0$ according to (24), which means that generically the boundary is a cylinder $M \times R$ with arbitrary curved spatial $M$. The problem with (23) and (24) is to prove that the equations are equations of the propagating type. This question will be considered elsewhere. Note that one cannot take the direct limit of the flat space-time by putting $g_{\mu\nu} = \eta_{\mu\nu}$ in (23). One has to consider $g_{\mu\nu} = \eta_{\mu\nu} + \kappa h_{\mu\nu}$ and retain the first order terms in $\kappa$ in $R'$. This just means that gravitons also contribute to the force on the boundary.

Note that if $T_{\mu\nu}n^\mu n^\nu > 0$ is negative then $R'$ is negative. That means that the boundary whose spatial curvature is everywhere negative, i.e. that looks, for example, like a distorted hyperboloid without "dents", tends to expand with non-zero acceleration. This fact suggests that space-times with positive "pressure" may be unstable with respect to the spontaneous internal boundary formation.

## 3. Remarks on Quantization

The formalism presented here was developed for the description of the boundary dynamics under the influence of the Casimir forces. It is hoped that two spontaneously created boundaries would like to develop an infinite area of the surface of their contact when they collide due to their expansion. This would produce an effective Kaluza-Klein compactification for the region between the two boundaries. It is known that a spherical ideal conductor boundary tends to expand infinitely while two parallel conductor planes tend to reduce distance between them to zero [5]. So the scenario described above is not that far-fetched.

However, the naive substitution of $(T_{\mu\nu})_{ren}$ instead of $T_{\mu\nu}$ in the boundary equation of motion (14) or (27) does not make much sense, since it is well-known that generically $(T_{\mu\nu})_{ren}$ diverges on the boundary. In deed, for the canonical scalar field (15), according to [4], $(T_{\mu\nu})_{ren}$ has the following asymptotic form in four dimensions and for Minkowski space-time

$$(T_{\mu\nu})_{ren}\big|_{\varepsilon\to 0} \propto \varepsilon^{-4}\left(-\frac{1}{16\pi^2}h_{\mu\nu}\right) + O(\varepsilon^{-3}), \tag{25}$$

where the Neumann boundary condition is assumed and with $\varepsilon$ being the distance to the boundary. The normal-normal component of $(T_{\mu\nu})_{ren}$ is less divergent.

$$n^\mu n^\nu (T_{\mu\nu})_{ren}\big|_{\varepsilon\to 0} \propto \varepsilon^{-3}\left(-\frac{1}{48\pi^2}\chi\right) + O(\varepsilon^{-2}), \tag{26}$$

We find that most of the divergent terms are insignificant for the boundary dynamics while the remaining divergent terms have the same functional form as the terms of boundary equation of motion (21). Thus there is hope that, when properly renormalized, Eq. (21) will somehow absorb at least some of the infinities in (26). The naive substitution of $(T_{\mu\nu})_{ren}$ in (23) might not be justified. Indeed on one hand we expect the condition $n^\mu T_{\mu\nu} h^\nu_\rho = 0$ to hold, together with $D^\mu T_{\mu\nu} = 0$, even after renormalization, since both are due to the coordinate invariance of the matter-gravity-boundary system. On the other hand, substitution o f (25) into (11) yields a generally non-vanishing divergent expression.

$$n^\mu (T_{\mu\nu})_{ren} h^\nu_\rho\big|_{\varepsilon\to 0} \propto \varepsilon^{-2}\left(-\frac{3}{80\pi^2}h^\mu_\rho D_\mu \chi\right) + O(\varepsilon^{-1}). \tag{27}$$

In general this is not zero. Taken literally, (27) means a breakdown of the coordinate invariance of the quantized system. This is unacceptable on general grounds. One possible explanation for (27) is that quantum fluctuations of the boundary must be taken into account to ensure that the right hand side o f (27) is zero.

Obviously more work on quantization of the interacting matter-gravity-boundary system is needed before one can have a definite answer to whether or not one can describe the motion of the boundary under the "pressure" from the Casimir "forces". If, however, one can find a consistent quantum formalism it may help resolving the mystery of spontaneous compactification in Kaluza-Klein theories by providing another possible mechanism for its dynamical realization, in addition to that currently discussed in the literature [7, 8].

## Acknowledgements

I am greatly indebted to Jeffrey Goldstone for numerous discussions on the subject. I wish to thank Alan Guth for valuable suggestions.